# Stochastic mechanisms forming large clones during colonisation of new areas

M. RAFAJLOVIĆ[1,2,†], D. KLEINHANS[2,3,¤,†], C. GULLIKSSON[4], J. FRIES[4], D. JOHANSSON[2,5], A. ARDEHED[2,5], L. SUNDQVIST[2,3], R. T. PEREYRA[2,6], B. MEHLIG[1,2], P. R. JONSSON[2,6] & K. JOHANNESSON[2,6]

[1] Department of Physics, University of Gothenburg, Gothenburg, Sweden

[2] The Linnaeus Centre for Marine Evolutionary Biology, University of Gothenburg, Gothenburg, Sweden

[3] Department of Biological and Environmental Sciences, University of Gothenburg, Gothenburg, Sweden

[4] Department of Applied Physics, Chalmers University of Technology, Gothenburg, Sweden

[5] Department of Biological and Environmental Sciences, University of Gothenburg, Tjärnö, Strömstad, Sweden

[6] Department of Marine Sciences, University of Gothenburg, Tjärnö, Strömstad, Sweden

[†] These authors contributed equally to this work

[¤] Current address: Department of Physics, University of Oldenburg, Oldenburg, Germany

Correspondence: Marina Rafajlović, Department of Physics, University of Gothenburg, Gothenburg, Sweden. Tel.: 004631786 9120; email:

marina.rafajlovic@physics.gu.se

Short running title: Dynamics of dominant clones

# Abstract

In species reproducing both sexually and asexually clones are often more common in recently established populations. Earlier studies have suggested that this pattern arises from natural selection favouring asexual recruitment in young populations. Alternatively, as we show here, this pattern may result from stochastic processes during species' range expansions. We model a dioecious species expanding into a new area in which all individuals are capable of both sexual and asexual reproduction, and all individuals have equal survival rates and dispersal distances. Even under conditions that eventually favour sexual recruitment, colonisation starts with an asexual wave. Long after colonisation is completed, a sexual wave erodes clonal dominance. If individuals reproduce more than one season, and with only local dispersal, a few large clones typically dominate for thousands of reproductive seasons. Adding occasional long-distance dispersal, more dominant clones emerge, but they persist for a shorter period of time. The general mechanism involved is simple: edge effects at the expansion front favour asexual (uniparental) recruitment where potential mates are rare. Specifically, our stochastic model makes detailed predictions different from a selection model, and comparing these with empirical data from a postglacially established seaweed species (*Fucus radicans*) shows that in this case a stochastic mechanism is strongly supported.

# Introduction

Most species in northern Europe, Asia and America established their present distributions following the last glacial maximum about 17,000 years ago. Currently, various human activities bring species into new territories, and this trend is expected to increase rapidly (Liu *et al*., 2006; Bax *et al*., 2003; Molnar *et al*., 2008; Franks *et al.,* 2014). Thus for a large number of species it is relevant to ask: what happens to species during and after an expansion into a new territory? Many studies focus on the effects of altered selection pressures (e.g. Keane & Crawley, 2002; Buckley & Briddle, 2014; Svenning *et al.,* 2014) while there is surprisingly little work examining the role of stochastic mechanisms. During the expansion, stochastic processes may be much more prominent than otherwise and affect the genetic structure of species over extended periods of time (Ibrahim *et al*., 1996; Excoffier *et al*., 2009). For example, experimental studies using unicellular and completely asexual organisms that expand into new areas show that populations that are originally composed of a homogeneous mix of genotypes turn into highly structured populations with dominance of single genotypes over large areas solely due to sampling effects (Hallatschek *et al*., 2007; Excoffier & Ray, 2008).

Many species have the potential to alternate or simultaneously reproduce both sexually and asexually (e.g. Schön *et al*., 2009; Vallejo-Marin *et al*., 2010; Dahl *et al*., 2012; Reichel *et al*., 2016). Among these, the incidence of asexual recruitment is commonly higher in new territories and in young habitats (Ting & Geller, 2000; Eckert, 2002; Kearney, 2003; Kliber & Eckert, 2005; Tatarenkov *et al*., 2005; Liu *et al*., 2006; Kawecki, 2008; Silvertown, 2008; Vrijenhoek & Parker, 2009). Earlier studies suggest this bias is due to selection favouring specific genotypes that are generally, or locally, successful in the new environment (Parker *et al*., 1977; Vrijenhoek, 1984; Peck *et al*., 1998; Stenberg *et al*., 2003; Kearney, 2005; Hörandl, 2009). Other studies suggest that selection removes the sexual function in order to prevent inbreeding in small marginal populations (Haag & Ebert, 2004; Pujol *et al*., 2009). Both groups of models have, however, difficulties in explaining why a majority of species with asexual reproduction has preserved sexual function in the absence of stabilising selection maintaining sexual traits (Green & Noakes, 1995;

Johnson *et al*., 2010).

Alternatively, and as suggested here, increased asexual recruitment in colonised territories may be explained by stochastic mechanisms with prominent effects on both the sex ratio and the genetic structure of populations. Asexual reproduction is uniparental, and uniparental reproduction generally provides a higher reproductive assurance than bi-parental reproduction when individuals have low mobility, low dispersal capacity of gametes, and live thinly scattered (Tomlinson, 1966; Pannell & Barrett, 1998). During establishment of species in new territories, populations at the front of the expansion have low densities, and asexual reproduction is in this case likely to be more efficient than sexual reproduction, unless selfing is possible. For similar reasons, uniparental reproduction is favourable if new habitats are reached by long-distance dispersal (Baker, 1955; Pannell & Barrett, 1998; Bialozyt *et al*., 2006).

The importance of increased asexual (uniparental) reproduction during colonisation of new territory in species that may not self-fertilise is poorly understood. Indeed, it is unclear how the spatial and temporal population genetic structure evolves over time if colonisation of new areas is by a combination of asexual and sexual recruitment. In large populations, very small proportions of sexual recruitment lead to similar equilibrium genotype diversity as for fully sexually recruiting populations (Bengtsson, 2003). However, during the transient phase of expansion into new areas, when populations have low densities and are fragmented, stochastic effects are likely to perturb genotype distribution away from equilibrium. This raises important and intriguing questions. For example, how does the spatial genotype structure of such an expanding population evolve in space and time? Are there significant differences between genotype structure in a transient as compared to an equilibrium state? For how long will differences last? And how do these effects depend on costs of sexual and asexual reproduction, dispersal type and rate, and life-history of species?

To answer these questions we use a spatially explicit model and analyse colonisation of a new territory by a species with separate sexes (dioecious), and both sexual and asexual reproduction. Most importantly, our model is fully neutral with respect to genotype survival, and as such contrasts with earlier models in which selection is assumed to act on genotype variation (e.g. Parker *et al*., 1977; Vrijenhoek, 1984; Peck *et al*., 1998). We have primarily designed the model to investigate a stochastic explanation to the strong spatial genetic structure of a young dioecious marine macroalga, *Fucus radicans*, that recently evolved and expanded its distribution in the Baltic Sea (Pereyra *et al*., 2009). The main characteristics of the geographic pattern of sex and clonal bias observed in this species is similar to what is found in several other species of plants, macroalgae and invertebrates (Ting & Geller, 2000; Eckert, 2002; Saltonstall, 2002; Kliber & Eckert, 2005; Liu *et al*., 2006; Kawecki, 2008; Silvertown, 2008; Vrijenhoek & Parker, 2009; Miglietta & Lessios, 2009; Halling *et al.,* 2013).

Most of the species that show increased asexual reproduction in recently colonised territories share histories of recent range expansion, a capacity of both sexual and asexual reproduction, no self-fertilisation, and a dominance of local recruitment. We investigate under which conditions a dominance of asexual recruitment arises at the expansion edge and for how long it typically persists also when asexual reproduction is more costly than sexual reproduction. Notably, our model is stochastic and does not involve selection among genotypes at any stage. Using this modelling approach, we ask how the spatial pattern of genotype variation depends on the underlying sexual structure evolving during colonisation. Specifically, we study how it depends on time since the start of expansion, population dispersal capacity, type of dispersed propagules (fragments or seeds), and mean lifetime of individuals. Finally, we compare a number of qualitative predictions of our stochastic model with the corresponding predictions of a selection-based model, to infer the relative importance of the stochastic and selection-based mechanisms for establishing a clonal-biased pattern, as, for example, observed in the seaweed *Fucus radicans*.

## Model and methods

**The biology and spatial genetic structure of *Fucus radicans***

The seaweed *Fucus radicans* formed less than 8000 years ago from *Fucus vesiculosus* inside the Baltic Sea (Pereyra *et al*., 2009). It has since spread and established along the coasts of the northern Baltic Sea (Bothnian Sea), and along the coast of Estonia in the Baltic Proper (Pereyra *et al.*, 2013). *Fucus radicans* is a perennial, dioecious, facultative asexual species with seasonal reproduction during the summer. The lifetime of an individual (the diploid thallus) is perceived as relatively long (~10 years, on average, see Discussion). After having reached maturity, both males and females produce gametes as well as asexual fragments (Tatarenkov *et al.*, 2005). Gametes are shed into the water and fertilised externally. The fertilised eggs settle close to the female because of their negative buoyancy (Serrão *et al.*, 1997). Fragments detach from the thallus and reattach to rocks and stones by formation of rhizoids (Tatarenkov *et al.*, 2005). There is no experimental data on how far from the parental plant a fragment usually lands, but it has negative buoyancy and is likely to land close to the parental individual in most cases. However, in very turbulent water (e.g. during storms) fragments, as well as larger pieces of thalli are occasionally transported long distances (as has been observed, see Fig. 1). Notably, a new individual grown from a fragment is functionally indistinguishable to one grown from a fertilised egg, that is, both produce gametes and asexual fragments.

*Fucus radicans* shows a remarkably complex genetic structure. In some areas of its distribution populations are almost completely sexually recruited, while in other areas populations are dominated by single clones (Johannesson *et al.*, 2011; Ardehed *et al.*, 2015). Variation among populations within the same region may be large, but the overall trend is a strong dominance of sexual recruitment in southeast and dominance of a few large clones in northwest of the species' distribution (Fig. 1). One notable observation is the presence of a large female clone that is dominant over more than 550 km of the Swedish coast (Fig. 1, blue sector). This may be due to this clone having superior fitness to other genotypes. Alternatively, stochastic demographic processes during the range expansion of *F. radicans* may have contributed significantly to the heavily biased spatial distribution of clonal genotypes and made this clone dominant. To test this hypothesis, we formulate and analyse a mathematical model.

**Model**

We assume a dioecious species colonising a new habitat in which each individual may reproduce both sexually and asexually. We do not take into account any adaptation to the conditions of the environment, that is, all genotypes have the same fitness. In this respect our approach offers an alternative to existing models analysing establishment in new habitats (e.g. postglacial, temporal or ecologically marginal environments) as these models assume selection among genotypes (see e.g. Peck *et al*., 1998; Hörandl, 2009).

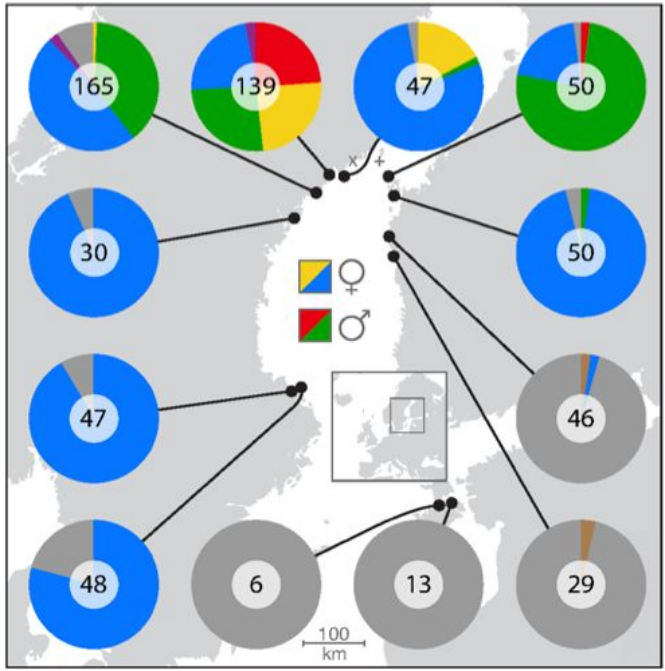

Figure 1. Distribution of *Fucus radicans* in the Baltic Sea, northern Europe. Grey shows unique individuals or local clones. Remaining colours denote clones that were found at more than one site. Each site is additionally labelled by the number of individuals sampled in it. The symbols "x" and "+" indicate locations where a free-floating individual (author obs.) and one single attached individual of *Fucus radicans* (reported in Albertsson and Bergström, 2008) were found, supporting occasional long-range dispersal outside the current known distribution of the species. The remaining data are taken from Ardehed et al. (2015). The genotype of the yellow clone suggests it is descended from a sexual cross between the blue female and the green male clones (see Ardehed et al. 2015).

Our model is constructed as follows (see also Supporting Information). A habitat consists of *N* patches of equal quality. Each patch can either be empty or occupied by one individual of known sex. In the model the patches are arranged linearly in a circle with periodic boundary conditions. This spatial configuration of patches bears a resemblance to the colonisation of a coastal region, but it is also relevant for populations distributed in a watershed of interconnected rivers and lakes, a high-altitude area surrounding an alpine landscape, or the peripheral parts of a desert.

In the model individuals have either one (annuals/semelparous) or multiple (perennials/iteroparous) reproductive seasons during their lifetime. In the beginning of each reproductive season, all adults disperse gametes and asexual propagules locally to their near-by sites (short-range dispersal). For simplicity, the short-range dispersal capability of gametes is assumed to be sex-independent and equal to that of asexual propagules. The short-range dispersal displacement $\Delta$ takes integer values from $-N/2$ to $N/2$, and it is sampled from a Gaussian distribution that is symmetric around zero. The distribution is described by a parameter $\alpha$ such that the mean short-range dispersal distance in units of habitat patches is roughly proportional to $\alpha$. In particular, the probability $K_S(\Delta)$ that an individual disperses a gamete or an asexual propagule by a displacement $\Delta$ via the short-range dispersal is given by

$$K_S(\Delta) = \frac{1}{\mathcal{N}_S} \sum_{k=-\infty}^{\infty} \exp\left[-\frac{(\Delta + k\,N)^2}{\alpha^2}\right], \text{ where } \Delta = -\frac{N}{2}, \ldots, \frac{N}{2}. \qquad (1)$$

Here the coefficient $k$ takes integer values from $-\infty$ to $\infty$. Note that this coefficient is introduced for mathematical reasons, due to periodic boundary conditions in the circular habitat with finitely many patches (because, mathematically speaking, such a circular habitat is infinitely long). However, for the values of the parameter $\alpha$ used in this study (see below), the terms with $k \neq 0$ in Eq. (1) are negligible compared to the terms with $k = 0$. Furthermore,

$$\mathcal{N}_S = \sum_{\Delta=-\frac{N}{2}}^{\frac{N}{2}} \sum_{k=-\infty}^{\infty} \exp\left[-\frac{(\Delta + k\,N)^2}{\alpha^2}\right] \qquad (2)$$

is a normalisation factor. We set the per-season per-individual rate of production of fertile eggs (for females), and sperms (for males) that survive dispersal to two. Note that this is not the total number of gametes produced by an individual but rather it corresponds to having one sexual recruit per individual in a well-mixed population with equal numbers of males and females (see below). Similarly, the per-season per-individual rate of production of viable asexual propagules that survive dispersal is set to a constant denoted by $c$. As two gametes of the opposite type are needed for each successful sexual recruitment, $c = 1$ means that asexual reproduction has the exact same potential for recruiting a new individual as sexual reproduction in a well-mixed population with equal number of males and females. Having this in mind, we hereafter refer to $c$ as '*cost of sexual as compared to asexual reproduction*'. When $c < 1$, asexual reproduction is more costly than sexual reproduction, and the opposite is true for $c > 1$. In this study, the main focus is on situations were $c < 1$.

Given the dispersal capabilities of propagules, and their rates of production, we compute the birth rates of sexual and asexual recruits in patch $i = 1, ..., N$ in a given reproductive season as follows. First, since sexual reproduction requires the presence of both male and female gametes in the same patch, we estimate the total contributions $p_f(i)$ and $p_m(i)$ to patch $i$ from dispersals by females and males, respectively. In our model, $p_f(i)$ is given by

$$p_f(i) = \sum_{j=1}^{N} \begin{cases} K_S[\Delta(i,j)], & \text{if individual in patch } j \text{ is female,} \\ 0, & \text{if individual in patch } j \text{ is male,} \\ 0, & \text{if patch } j \text{ is empty.} \end{cases} \quad (3)$$

Here $\Delta(i,j)$ is the displacement between patches $i$ and $j$. The corresponding contribution $p_m(i)$ from dispersals by males to this patch is obtained by replacing "male" by "female", and vice versa on the right-hand side in Eq. (3). Thus, females and males contribute $cp_f(i)$ and $cp_m(i)$ asexual propagules, and $2p_f(i)$ and $2p_m(i)$ gametes to patch $i$. Second, we assume that the birth rate $b_c(i)$ of asexual recruits in patch $i$ is

$$b_c(i) = c[p_f(i) + p_m(i)]. \quad (4)$$

Third, the birth rate $b_s(i)$ of sexual recruits in patch $i$ is assumed to be limited by the minimum of the total number of potentially successful gametes (as described above) contributed by females, and those contributed by males to this patch, namely

$$b_s(i) = \min[2p_f(i), 2p_m(i)]. \quad (5)$$

As already noted, when $c = 1$ and $p_f(i) = p_m(i) = 1/2$, the potential for sexual reproduction is equal to the potential for asexual reproduction in patch $i$.

After reproduction adults die at a per-season per-individual death rate $d$, that is, individuals have on average $d^{-1}$ reproductive seasons in their lifetime. When $d = 1$, all adult individuals die after the first reproductive season (an annual life cycle). Otherwise, when $d < 1$, an individual on average experiences multiple reproductive seasons, and the model describes a perennial life cycle. An adult that dies empties its patch. Each empty patch can be occupied by a single recruit sampled randomly from the sexual and asexual recruits produced in this patch. We assign to a clonal offspring the sex of its parent, whereas sexual recruits are males or females with equal probability. At the end of the reproductive season, the new recruits that establish in the habitat are treated as reproductively mature adults.

Finally, prior to the next dispersal of propagules, the adults or vegetative fragments may relocate through the process of long-range dispersal at a per-individual per-season rate $r$. When $r = 0$, long-range dispersal does not occur. Otherwise, when $r > 0$, long-range dispersal occurs, and we assume that it occurs rarely ($r \ll 1$). The destination patch is drawn from a symmetric power-law distribution with a parameter $\beta$ (Supporting Information). When $\beta$ is larger than, but sufficiently close to unity, typical long-range dispersal distances are of the order of the size of the available habitat. If the destination patch is already occupied, the dispersing individual dies. Otherwise, the dispersing individual occupies the destination patch. Note that rare long-range dispersal of clonal fragments, asexual offspring or apomictic seeds (hereafter collectively referred to as 'asexual propagules') are valid alternatives to rare long-range dispersal of adult individuals in the model (see Supporting Information).

In addition to the model described, we also analyse two modified versions. In one (see Supporting Information), we assume that only females have the capacity for both sexual and asexual reproduction, whereas males are not able to reproduce asexually (all else being the same as in the model explained above). In this model, the birth rate of clonal recruits in patch $i$ is computed as

$$b_c(i) = cp_f(i), \tag{6}$$

whereas the birth rate of sexual recruits is computed according to Eq. (5). In another modified version, we assume long-range dispersal of sexually produced seeds instead of adults or asexual propagules. For further details, see Supporting Information.

**Parameter choices**

We investigate the models for different combinations of parameter values (Table 1). The values of the death rate $d$ are chosen from a wide range to include annual species ($d = 1$) up to very long-lived perennial species ($d = 0.005$, 200 reproductive seasons). For the clonal birth rate ($c$) we mostly use $c \leq 1$ implying that asexual reproduction is either equally or more costly than sexual reproduction, but we also consider some cases when asexual reproduction is less costly than sexual reproduction. In most of the simulations we set the habitat size (total number of patches) to $N = 4000$, but we also analyse the model for $N = 2000, 8000,$ and 16000. To obtain a mean short-range dispersal distance that is much smaller than the habitat size we set $\alpha = 3$, or 6. Conversely, to accomplish long dispersal distances in the process of long-range dispersal, we set the parameter $\beta$ to a value close to unity ($\beta = 1.25$). For long-range dispersal to be rare, we choose small values of the long-range dispersal rate ($r \ll 1$). Most of the simulations are run for up to 20,000 generations.

| Parameter | Explanation | Values |
|---|---|---|
| $N$ | Number of patches in the habitat | 2000, 4000, 8000, 16000 |
| $c$ | Per-season per-individual clonal birth rate | 0.02, 0.2, 0.5, 0.6, 1, 1.2, 2 |
| $d$ | Per-season per-individual death rate | 0.005, 0.01, 0.1, 1 |
| $\alpha$ | Short-range dispersal parameter | 3, 6 |
| $\beta$ | Long-range dispersal parameter | 1.25 |
| $r$ | Per-season per-individual long-range dispersal rate | $0, 10^{-4}, 10^{-3}$ |

**Table 1.** Parameters of the model, their explanations, and the values used in the computer simulations.

To be able to characterise the genetic structure of the population we label each individual at the start of each simulation with a unique number ('ID') that is copied upon long-range dispersal and asexual reproduction. Individuals originating from sexual reproduction receive a new unique ID. Simulations are initialised with 100 populated patches arranged side by side around a single origin, neighbouring individuals having opposite sexes (50 females and 50 males). We specifically choose this initial condition because it assures maximum mixing of males and females at the start of the expansion (alternating sexes), has no sex bias at small local scales and hence minimises the risk that asexual reproduction is initially promoted over sexual. This starting arrangement mimics a population that has survived through a perturbation (e.g. a glacial period) in a small area and from there expands into an empty habitat when conditions improve. Alternatively, it mimics a population that expands into a new territory after a human introduction to one part of the new area.

Finally, because the model is stochastic, for each set of parameters we perform 100 independent simulations (unless stated otherwise).

## Results

Due to the effect of random fluctuations, all populations bounded by a finite maximum size and with finite per-individual birth rates must eventually experience extinction (Eriksson et al. 2013). For the model parameters tested (Table 1) we observe that the population initialised with 100 individuals experiences rapid extinction when individual death rate is high ($d = 1$, corresponding to annuals), the mean short-range dispersal distance is small ($\alpha = 3$, or 6), and the clonal birth rate is low ($c \leq 0.2$). These cases are, therefore, omitted from further analysis. Furthermore, due to random sequences of births and deaths in a population of finite size, the population must eventually reach the state where one sex is globally lost and the other is fixed (unless the population experiences extinction prior to global fixation). The time to reach global fixation is expected to scale exponentially with the population size. Within the tested values of the model parameters (excluding the parameters for which the population experiences rapid extinction), we do not find any cases of global fixation, except under the modified version of the model in which males are incapable of asexual reproduction (see Discussion and Supporting Information). Therefore global fixation is not further considered here.

**Short-range dispersal dynamics**

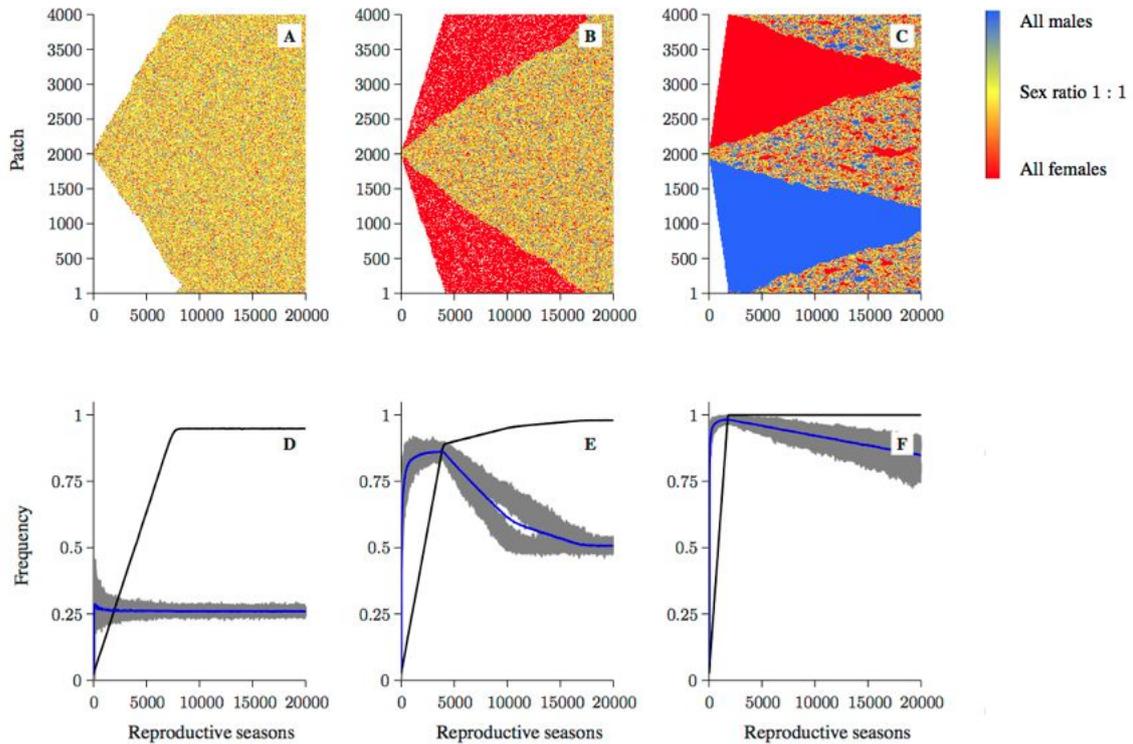

Figure 2. Sexual patterns in populations with different costs of sexual as compared to clonal reproduction ($c$). Panels A-C: space-time patterns of local sex ratios (pools of 8 neighbouring patches) obtained from single stochastic realisations of the model with $c = 0.2$ in A, $c = 0.5$ in B, and $c = 1$ in C. Empty patches are coloured white. Panels D-F: time dependence of the average frequency of occupied patches (black line), and of the average frequency of asexually recruited individuals (blue line) for the parameter values in A-C, respectively. Grey lines depict the results of independent simulations (100 runs). Note that, due to periodic boundary conditions, the decrease of asexual recruitment after colonisation is faster when the clonal colonies established at the different sides of the origin of colonisation are of the opposite sexes (grey lines below the solid blue line in E, F) than when they are of the same sex (grey lines above the solid blue line in E, F). Initial configuration: 100 neighbouring patches occupied, alternating sexes. Remaining parameters: $N = 4000$, $\alpha = 3$, $d = 0.1$, $r = 0$.

When long-range dispersal is not possible ($r = 0$), the sexual structure and the underlying genetic structure of the population depend sensitively on the cost of sexual as compared to asexual reproduction (i.e. $c$), as well as on the death rate ($d$). When the cost of clonal as compared to sexual reproduction is high ($c \ll 1$), the sexes are essentially homogeneously distributed both in space and time, with only small patches of single-sex dominance (Fig. 2A, D). By contrast, for lower costs of clonal reproduction ($c > 0.2$ for $d = 0.1$), large single-sex (and single-genotype) colonies are established, one at each edge of the expansion (Fig. 2B-C, E-F). Note, however, that rate of clonal reproduction need not be larger than rate of sexual reproduction for this to happen. Similarly, with a high death rate (short life-span), an essentially homogeneous distribution of sexes appears in space and time (Fig. 3A, D, S1A), whereas large single-sex colonies are formed for perennial species with longer life-times (Fig. 3B-C, E-F).

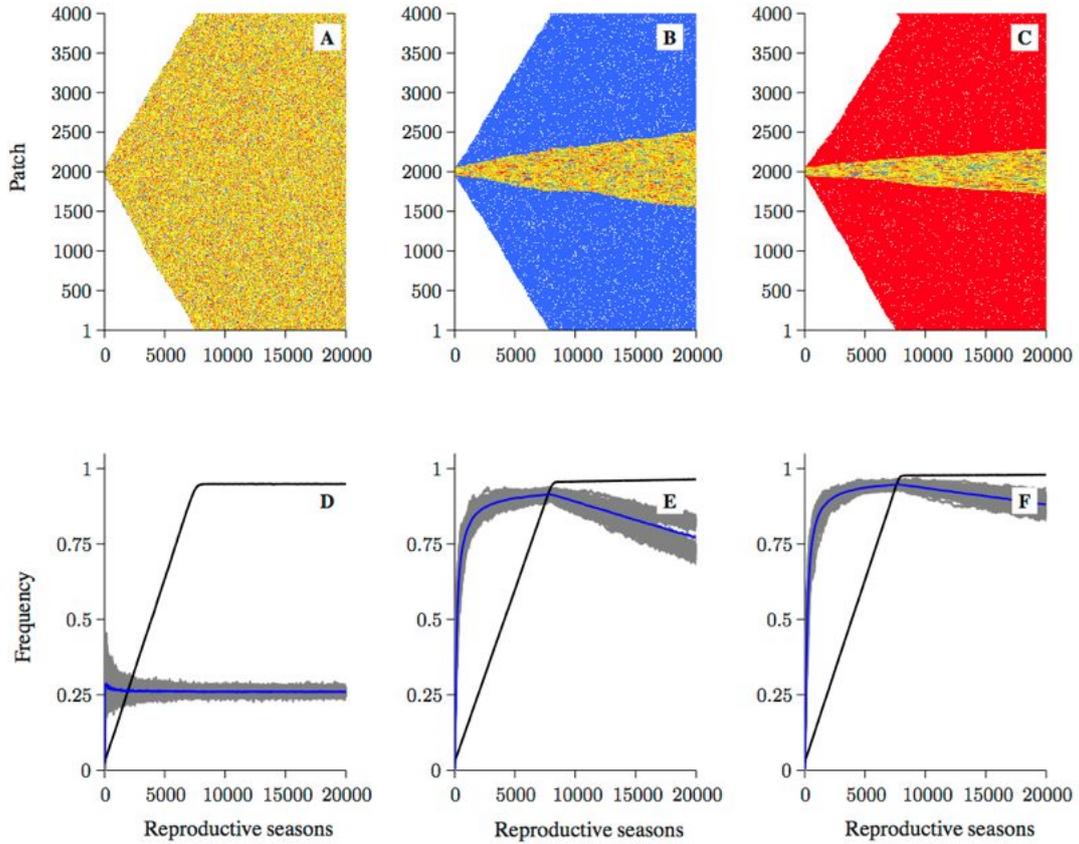

Figure 3. Sexual patterns in populations with different death rates $d$. The values of the death rate are: $d = 0.1$ in A and D, $d = 0.01$ in B and E, and $d = 0.005$ in C and F. For the explanation of the results shown in different panels, refer to the caption of Fig. 2. The results in panels A, and D correspond to those in Fig. 2A, and Fig. 2D, respectively. Remaining parameters: $N = 4000, \alpha = 3, r = 0, c = 0.2$, 100 stochastic realisations in D-F.

In cases of clone formation, we observe three phases of spatial distribution of sexes. The first phase occurs during the spread into the new territory and is characterised by an expansion of single sex-colonies consisting of single clones. The "asexual wave" moves faster for higher rates of clonal reproduction (Fig. 2B-C). While sexual recruitment of new individuals is prevented in the single-sex areas, sexual reproduction progressively spreads from the centre of the expansion ("a sexual wave", see spread of regions with a homogeneous sex ratio in Figs. 2-3). The speed of the sexual wave is slower for higher cost of sexual as compared to clonal recruitment (Fig. 2A-C) and for longer life-spans (Fig. 3A-C). Notably, for the parameters set in Fig. 2B (e.g. average life-time of 10 years) the population is dominated by a pair of large clones during colonisation and long thereafter (12,000-18,000 years, Fig. 2E) despite, in this case, the cost of asexual reproduction being twice the cost of sexual reproduction ($c = 0.5$).

The second phase starts roughly when the habitat becomes fully occupied. In this phase large single sex colonies formed during colonisation shrink in size as the region where sexual reproduction is possible expands. The sexual wave expands slowly and clonal colonies persist for longer times under higher costs of sexual as compared to clonal reproduction and/or longer life-span of individuals (Figs. 2E-F, 3E-F).

The third phase starts when the dominant clonal colonies finally disappear. In this phase the overall frequency of asexually recruited individuals fluctuates around a constant value suggesting that the population has reached a quasi-steady state. ("Quasi" because, as pointed out above, the population must eventually experience global fixation, unless it experiences extinction first.) The frequency of asexually recruited individuals in this phase is larger when the clonal birth rate $c$ is higher (compare Fig. 2D to Fig. 2E at the end of the timespan shown), but it is approximately the same for different values of death rate $d$ (results not shown). Note also that the sexes in this phase are essentially homogeneously distributed and single-sex colonies with up to roughly a hundred individuals appear temporarily (e.g. Fig. 2C). In contrast to the single-clone colonies that dominate during colonisation, each single-sex colony in the quasi-steady state typically consists of multiple clones of the same sex. Finally, single-sex colonies and the overall frequency of asexually recruited individuals in the three phases discussed are larger for smaller values of short-range dispersal (smaller $\alpha$) (compare Fig. 2B and E to Fig. S2A and D).

**Dynamics with occasional long-range dispersal**

The three phases of spatial distribution of sexes discussed above are also present in populations with occasional long-range dispersal, although with several noticeable differences to dynamics with only short-term dispersal. Firstly, long-range dispersal increases the speed of colonisation of the empty habitat (Fig. 4, S3). Secondly, with long-range dispersal a clone can spread and dominate in multiple (distant) areas (Fig. 4B-C). Thirdly, more than two dominant clones typically form during the colonisation phase with long-range dispersal (compare Fig. 4B-C, to A). The reason is that additional clones could be established following a long-range dispersal event of a fragment (or similar) of a genotype that is a sexual descendant of dominant clones.

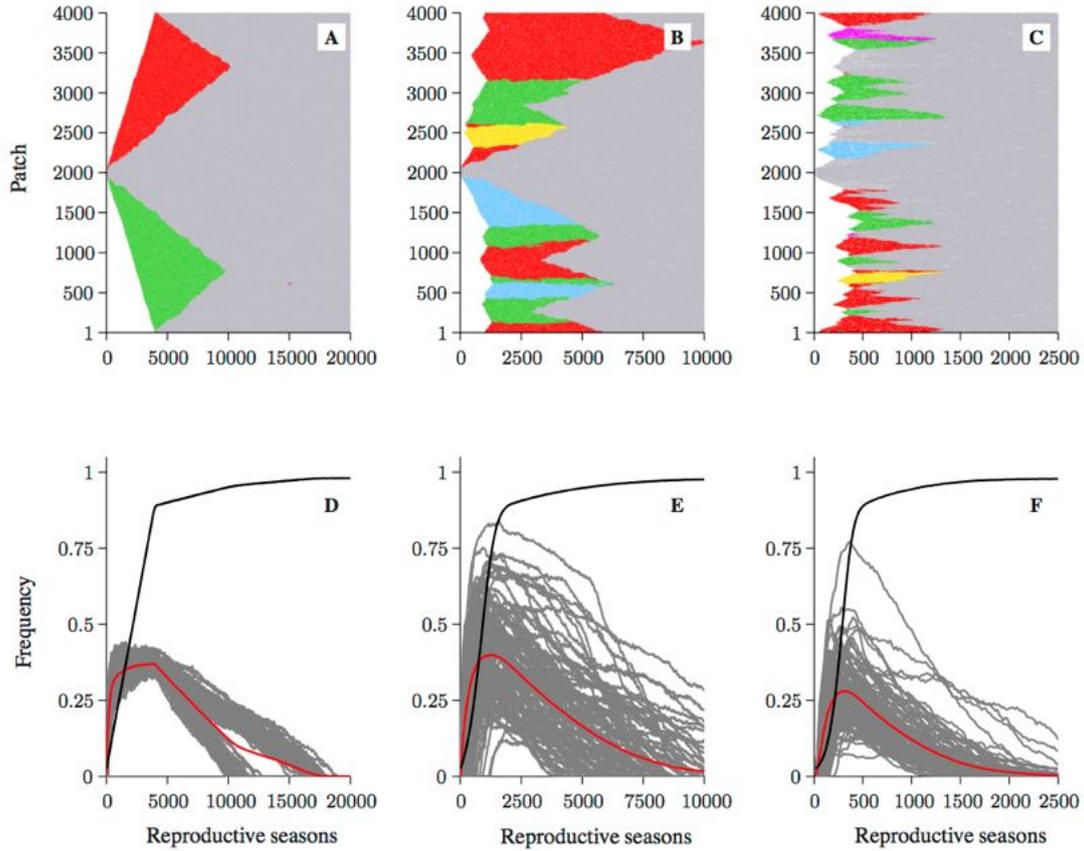

Figure 4. Dominant clones. Panels A-C: space-time patterns of the five largest clonal colonies (coloured red, green, blue, yellow and magenta) with only short-range dispersal in A ($r = 0$), and with both short-range dispersal and occasional long-range dispersal of adults, fragments or asexual propagules in B, C ($r = 10^{-4}$, and $r = 10^{-3}$, respectively). Genotypes differing from the five largest clones are coloured grey. Empty patches are coloured white. Panels D-F: grey lines show the frequency of the largest dominant clone as a function of time from 100 independent realisations of the model for the parameters used in panels A-C, respectively. Red lines are averages over the individual runs. Black line shows time dependence of the average frequency of occupied patches. Blue line shows time dependence of the average frequency of asexually recruited individuals. Remaining parameters used: $\beta = 1.25$, $N = 4000$, $\alpha = 3$, $d = 0.1$, $c = 0.5$, 100 stochastic realisations in D-F. The spatial sexual patterns corresponding to the dominant clone patterns shown in panels A-C are depicted in Fig. S3A-C, respectively.

Fourthly, the persistence time is on average shorter, and the size of individual clonal colonies formed during colonisation are on average smaller with long-range dispersal, and this effect is stronger for higher rates of dispersal (Fig. 4). Note, however, that the persistence time and size of the dominant clones are very different between individual runs with otherwise the same parameter values (grey lines in Fig. 4E-F).

In larger habitats the spatial patterns emerging in the model are similar to those described above (Figs. S4-S5), although large single-sex colonies, and widely spread dominant clones are more likely to be obtained in larger habitats (grey lines in Figs. S4D-F and S5D-F). As expected, the number of local colonies established by a dominant clone increases with increasing the habitat size (Fig. S5A-C). These effects arise because typical long-range dispersal distances are larger in larger habitats, and so are chances for a single clonal lineage to establish in multiple, distant locations.

**Dynamics in two alternative versions of the model**

In the model with males reproducing solely sexually, and females reproducing both sexually and clonally, we find that the main model predictions are retained, except that clonal colonies in this case consist exclusively of females (Fig. S6), and that for very high clonal birth rates ($c \geq 1$), males are likely to go extinct during the early colonisation phase (Fig. S6C).

Finally, in the model with long-range dispersal of sexually produced seeds instead of clonal fragments, we find that local clonal structures are likely to be formed (Fig. S7), but the geographic spread of a dominant clone is smaller than when clonal fragments have the potential to disperse long distances (Fig. S8). Consequently, with dispersal of sexual seeds it is very unlikely that a single clone dominates multiple distinct and widespread geographic areas.

# Discussion

Our model applies to dioecious species with predominantly local dispersal of recruits and a capacity for both sexual and asexual reproduction. Examples of such species are found frequently among seaweeds and invertebrates, and to a minor degree among aquatic and terrestrial plants. In what follows, we discuss our model results under the above mentioned assumptions of sexual systems and life-histories.

**General results and model mechanisms**

We know from earlier studies (Baker, 1955) that local clonal colonies are expected to form when a population with long-range dispersal capacity colonises a new area. Now our model results suggest that in perennial dioecious organisms, clonal dominance arises and persists over an extended period of time even without long-range dispersal. The clonal dominance during expansion results from single-clone colonies that compose the front of the colonisation ("an asexual wave"). Notably, an asexual wave will develop even when the cost of asexual reproduction is higher than the cost of sexual reproduction. If the cost of asexual reproduction is instead lower than the cost of sexual reproduction ($c > 1$) dominance of asexual reproduction is expected everywhere in the species' distribution, and not only during colonisation of a new habitat. (This result is obvious and not considered further here.)

An asexual wave formed during colonisation of a new habitat is a transient phenomenon, and occurs due to two effects. Firstly, stochastic demographic fluctuations at the expansion front lead to establishing, by chance, a few nearby individuals of the same sex. Secondly, a difficulty of finding a mate at the expansion front impedes sexual reproduction locally. These two effects together contribute to the formation of local populations with strongly biased sex ratios along the expansion front. As sexual recruits are simultaneously produced and spread from the central part of the population, unbiased sex ratios will at a later stage, by means of a "sexual wave", be retrieved in areas initially occupied by clones. Hence, after the colonisation of an empty habitat is completed, the asexual wave is brought to a halt and the sexual wave starts to catch up and progressively erode the clonal colonies. However, this process is slow when individuals are long-lived ($d$ is small), because the sexual wave can only progress when already occupied patches at the expansion front are emptied.

In fact, the speed of the sexual wave is roughly proportional to $d$ (compare slopes of blue curves in the post-colonisation phase in Fig. 3B-C).

For most combinations of death rate ($d$) and clonal birth rate ($c$) asexual reproduction at the expansion front dominates over sexual reproduction and forms clonal colonies (Fig. 2-3). This is true for highly perennial populations ($d \ll 1$) even when sexual recruitment is much less costly than asexual recruitment ($c \ll 1$), as well as for shorter-lived organisms (higher $d$) with cost of asexual recruitment approaching (or higher than) cost of sexual recruitment (i.e. $c$ closer to, or higher than unity). For annual species, circumstances under which an asexual wave is formed are much more restrictive (under the criteria that asexual reproduction remains more, or equally costly as sexual reproduction). This somewhat contrasts the expectations in a metapopulation scenario with low migration, and/or high extinction rates of local populations, as investigated by Pannell and Barrett (1998). In their study, uniparental (e.g. asexual) reproduction in annual species provided higher reproductive assurance than bi-parental (e.g. sexual) reproduction, while conditions were less restrict in perennial species. This finding, however, is based on a comparison of the capacity for colonisation of empty patches by annual or perennial bi-parental species, as compared to uniparental species. In this case, perennial bi-parental immigrants benefit from a longer waiting time to find mates and establish colonies, and are thus more successful than annual bi-parental species. The model presented here, however, considers species with capacity for both asexual and sexual reproduction. As explained above, in this case colonisation typically occurs by means of an asexual wave. During the colonisation of a new habitat, the first wave (the asexual wave) more persistently prevents colonisation by the second wave (the sexual wave) if individuals are long-lived.

In our model, an annual population with a poor asexual capacity may fail in colonising a new habitat altogether. However, with individuals of a long lifetime, or a high per-season per-individual rate of production of sexually recruited offspring, risk of failure diminishes (results not shown), as also pointed out by Pannell and Barrett (1998) but for obligate sexual species.

Adding occasional long-range dispersal in our model to a species with mostly restricted dispersal has both short-term and long-term effects. Primarily it facilitates colonisation of new habitats by already established dominant clones so that they appear in multiple sites. In this way, and in line with what was originally suggested by Baker (1955), long-range dispersal increases the spatial separation of sexual and clonal recruits. By this mechanism, occasional long-range dispersal also increases the number of dominant clones. The reason for this is that new genotypes produced in areas with sexual activity occasionally disperse and establish in new areas and from here initiate new asexual waves. Secondarily, occasional long-range dispersal also promotes the introduction of individuals of opposite sex into established clonal areas, thereby facilitating the erosion of existing clonal colonies.

The general pattern of our model - the expansion by an asexual wave - is akin to earlier findings of "allelic surfing" (Edmonds *et al*., 2004; Klopfstein *et al*., 2006, Excoffier & Ray, 2008; see also a discussion on "area effects" in e.g. Goodhart, 1963). In both allelic surfing and under area effects, stochastic processes rather than selection determine the structure of populations during transient stages. These transient stages may be short or long. In our model, perennials of average life-times of 5-10 years, or more, form dominant clones that persist for thousands to tens of thousands of reproductive seasons (Fig. 3). This time scale is relevant not only for very recent species invasions but also for postglacial colonisation of northern habitats.

**Comparison of model results with empirical data**

In our model, the mechanism by which some genotypes come to dominate the new population is primarily stochastic. In this respect, this model offers an alternative, or a complement, to models that suggest structural patterns after colonisation to be a consequence of selection on genotype variation (e.g. Parker *et al*., 1977; Vrijenhoek, 1984; Peck *et al*., 1998; Kearney, 2005). To illustrate how the two types of models may be contrasted and compared, we first examine the predictions of our stochastic model if parameterised by data from the seaweed *Fucus radicans*. In the next step we outline predictions of our stochastic model and of the selection-based model, and compare qualitative patterns observed in *F. radicans* with model predictions.

As already described, *F. radicans* is dioecious and both females and males have capacity for sexual and asexual reproduction. Although it is known to be long-lived and perennial, to our knowledge there is no direct estimate of individual life-time in this species. We have instead used data from the related fucoid species (*Ascophyllum nodosum*) that lives in similar habitats as *F. radicans* in the Atlantic. Measuring survival rates of marked individuals Åberg (1992) estimated maximum life-time in *A. nodosum* to be 50-60 years. With this as a guideline we assume an average life-time in *F. radicans* of at least 10 years, corresponding to $d \leq 0.1$ in our model. Under this assumption, the minimum cost of sexual as compared to clonal reproduction ($c$) for which we expect clonal dominance at the expansion front in *F. radicans* is larger than 0.2 (Fig. 2). As we find large areas where clones are rare (e.g. Estonia, see Fig. 1) we assume that asexual reproduction is more or equally costly than sexual reproduction. Thus we conclude that $0.2 < c \leq 1$ in *F. radicans*. For $c = 0.5$ and $d = 0.1$, and with occasional long-distance dispersal (as observed in *F. radicans*, see Fig. 1), our stochastic model generates two asexual waves, one in each of the two expansion fronts, resulting in large areas dominated by single clones. After occasional long-distance dispersal, additional clones get established (Fig. 4). After about 1000 years (Fig. 4B) the new area is completely filled up, but it takes additionally 10,000 years before the sexual wave has finally eroded the large clones. Note that these estimates are based on our model results with $N = 4000$. In reality, however, the total number of patches (individuals) is likely much larger. In this case, keeping all other parameters the same, the time to the completion of colonisation, and the persistence time of clonal colonies is expected to be even larger than the estimates given above. We also note that the parameter $N$ used in our main results speeds up the simulations. The effects of a larger habitat size are assessed by additional simulations (Figs. S4-S5).

Five specific predictions from our stochastic model can be compared with empirical data retrieved from earlier studies of *F. radicans*.

*Prediction 1.* The stochastic model predicts a strong distributional bias in sex ratio in recently colonised areas dominated by one sex and single clones, while the sexual wave has eroded this bias near the original area. Under selection, asexual recruitment is expected in ecologically marginal habitats (Vrijenhoek, 1984; Peck *et al*., 1998). Unfortunately, these predictions are difficult to separate in practice, as the younger habitats may also be the more ecologically marginal habitats. This seems true in *F. radicans* where colonisation presumably started in central Baltic and proceeded towards younger and more marginal areas in the north. Model comparison thus remains inconclusive in this case.

*Prediction 2.* The stochastic model predicts that the few clones that happen to be at the front at start of the expansion will efficiently spread by the asexual wave and become widely distributed. Under selection it is more likely that different environmental conditions will favour different local clones recruited from local sexual events. In *F. radicans,* the largest clone has an extensive geographic distribution being dominant over 550 km of shore line spanning latitudinal gradients in, for example, salinity and temperature (Fig. 1). It seems unlikely that selection is uniform over the wide ecological and geographic ranges occupied by this large clone, and its distribution is more compatible with predictions of the stochastic model.

*Prediction 3*. Following the arguments above, the stochastic model predicts the most dominant clones to be the same as those established at the front of the asexual waves when these waves were initiated. The dominant clones are thus, most likely, the oldest clones. Under selection, by contrast, it seems likely that, over time, earlier dominant clones are replaced by new and more competitive ones introduced by long-range dispersal. In *F. radicans* the dominant clones are indeed very old. The most widespread clone, for example, has accumulated somatic mutations that suggests this clone may be thousands of years old (Ardehed *et al*., 2015). This observation is therefore, more compatible with a stochastic mechanism.

*Prediction 4.* Under selection, fitness of a locally dominant clone is expected to be higher than fitness of other genotypes present in the same area, while a stochastic model makes no such predictions. Empirical tests of genotype fitness are limited by what aspects of fitness are actually tested, and in what context. Keeping this in mind, experimental tests of performance of some dominant clones (including the most widely distributed clone) in *F. radicans* show that none of them are more tolerant to physical stresses than sexually recruited genotypes from the same area (Johannesson *et al*., 2012). Moreover, the most widely distributed clone has a similar growth rate and similar capacity for asexual reproduction as two less widely distributed clones (Johansson, 2013). This far, there is thus no support for a positive correlation between clonal dominance and fitness. This prediction is, therefore, also more compatible with a stochastic than a selection-based mechanism.

*Prediction 5*. Under a stochastic scenario there is no selection among genotypes with different capacities for asexual and sexual reproduction. Under selection, on the other hand, successful genotypes maintained by asexual reproduction will be favoured in ecologically marginal areas. As there will be a trade-off between investing in sexual and asexual propagules, sexual reproduction is likely to be locally lost, or reduced, by selection in marginal, as compared to central areas. In particular, old dominant clones are most likely to have lost (or significantly reduced) their sexual function. In *F. radicans* members of the large clones remain sexually active and produce gametes at similar rates as other individuals (Forslund & Kautsky, 2013), supporting a stochastic model over a selection-based model.

**Model relevance - other species**

Sex ratios are biased in many species that have invaded new territory (see Introduction), but even more interesting is that in several of these species a few large clones dominate the new area despite a mix of genotypes being present in the native area (Ting & Geller, 2000; Saltonstall, 2002; Eckert, 2002; Reitzel *et al.,* 2008; Halling *et al*., 2013). For some of these species it has also been observed that members of the dominant clones have the potential to be sexually active (Green & Noakes, 1995; Johnson *et al*., 2010; Kliber & Eckert, 2005). This suggests that the reason for cloning in these cases is not primarily selection against sexual recruitment.

We have modified our model to test its relevance for some additional life-history strategies. Thus for a dioecious species where only one sex is able to reproduce both sexually and asexually (e.g. water fleas and aphids, Yin *et al*., 2012; Gilabert *et al*., 2015) the model outcome does not change significantly. Our model of long-range dispersal of adults or vegetative parts is relevant for macroalgae, aquatic invertebrates, and some aquatic plants (Ingólfsson, 1995; Rothausler *et al.,* 2015) but many terrestrial plant species disperse long distances by sexually produced seeds. The model results show that this will lead to formation of local clones that are unlikely to spread to multiple distant areas. Empirical data support this model prediction as widely distributed dominant clones are found in species dispersed by adults or fragments living in aquatic environments (Freeland *et al*., 2000; Ting & Geller, 2000; Eckert *et al*., 2003; Kliber & Eckert, 2005; Darling *et al*., 2009; Pettay *et al*., 2009; Koenders *et al*., 2012), while aquatic and terrestrial plants dispersed by sexual seeds tend to have local clones (Waycott *et al*., 1996; Carino & Daehler, 1999; Reusch, 2001; Paun *et al*., 2006; Wilk *et al*., 2009).

**Conclusion**

A geographic sexual bias with large and dominant clones in some areas and mostly sexual recruitment in other areas can be generated under two different but not mutually exclusive mechanisms, namely (*i*) selection for fixation of the most fit genotype, and (*ii*) stochastic demography together with edge effects during range expansion (Fig. 5, right and left, respectively). In what follows we do not consider species in which asexual recruitment is overall more efficient than sexual recruitment, as such species are expected to be highly asexual over their complete distribution. Furthermore, we focus on a post-colonisation stage that is transient but long-lasting. We find that whether the prevailing mechanism for establishing a clonal bias is stochastic or selection-based depends on the individual's life-span and the relative cost of sexual and asexual reproduction. (The more detailed pattern depends on the rate of long-distance dispersal and if this is by adults, vegetative parts or seeds.) For species with a longer life-span, expansion is dominated by the asexual wave of colonisation. Consequently, genotype distribution is largely stochastic, even when cost of asexual reproduction is relatively high compared to cost of sexual reproduction (Fig. 5, left). Notably, under the asexual expansion wave, selection is efficiently eliminated by the absence of genotype variation. For an annual species, by contrast, stochastic mechanisms during expansion are weak, and instead it expands by a sexual wave (Fig. 5, right). Sexual reproduction is here a major source of new genotypes, and this provides an opportunity for selection for generally or locally fit genotypes that may be fixed under selection against sexual recombination. Indeed, selection has widely been invoked as a general mechanisms behind dominance of asexual recruitment in recently, and relatively recently (postglacial) invaded areas (Parker *et al*., 1977; Vrijenhoek, 1984; Peck *et al*., 1998; Kearney, 2005). Our conclusion, however, is that under circumstances of long-lived (perennial) dioecious and mainly locally recruited species, a fully stochastic model is more likely to explain clonal dominance and sexual bias in recently, or relatively recently colonised areas. The clonal dominance is transient, but even so it persists over time periods relevant for most human mediated and postglacial invasions. Eventually, following the sexual wave, the transient stage will be replaced by an equilibrium-like stage during which genotype variation will recover and selection may (or may not) favour fixed clones and again removal of sexual reproduction.

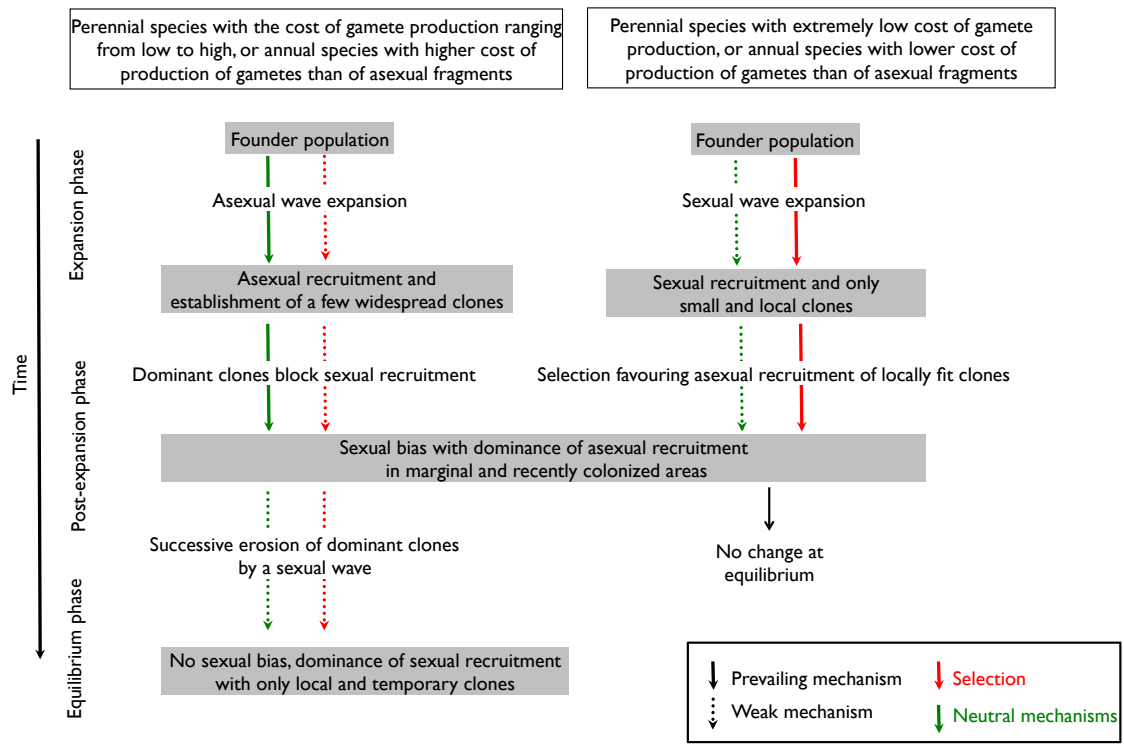

Figure 5. Schematic comparison of the expansion model with strong neutral stochastic processes and no or comparatively weak selection on individual genotypes (left), to a model with weak neutral stochastic processes and comparatively strong selection (right). Note that the absolute strength of selection may be the same in the two cases, but its strength relative to the corresponding strengths of neutral processes is different. Here, the term "neutral processes" refers to stochastic demographic fluctuations and the difficulty of finding a mate of the opposite sex at the expansion front.

As illustrated here for the seaweed *Fucus radicans*, a modelling approach can provide predictions that can be compared with empirical data on age and distribution of single clones, annual or perennial life-styles, estimates of average life-span, mechanisms and rate of dispersal, and (if available) relative cost of sexual and asexual reproduction. In conclusion, we find that such a qualitative comparison is necessary to argue for a stochastic or a selection-based explanation in individual species.

## Data accessibility

The computer codes used to simulate the model will be submitted to Dryad.

## Acknowledgements

Funding for this work was provided by the Centre of Marine Evolutionary Biology (www.cemeb.science.gu.se) at University of Gothenburg, supported by the Swedish Research Councils (Formas and Vetenskapsrådet) and from the EU BONUS program to the research program BONUS Bambi, as well as by additional grants from Vetenskapsrådet, and from the Göran Gustafsson Foundation for Research in Natural Sciences and Medicine. MR and BM were supported in part by VR grant no. 2013-3992. The authors thank the colleague members of the Centre for Evolutionary Biology at University of Gothenburg for supporting discussions during the preparation of this manuscript.

**Supplementary figures**

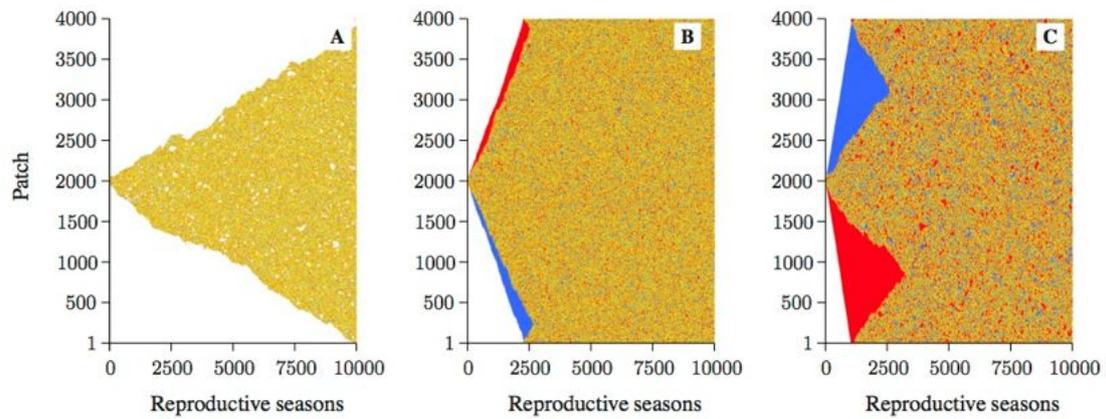

Figure S1. Spatial sexual patterns in annuals. The value of the death rate in all panels is $d = 1$. The panels differ by the cost of sexual as compared to clonal reproduction: $c = 0.6$ (costly to recruit clonal offspring) in A, $c = 1.2$ in B, and $c = 2$ (costly to recruit sexual offspring) in C. For the explanation of the results shown in the panels, refer to the caption of Fig. 2 in the main text. Remaining parameters: $N = 4000, \alpha = 3, r = 0$.

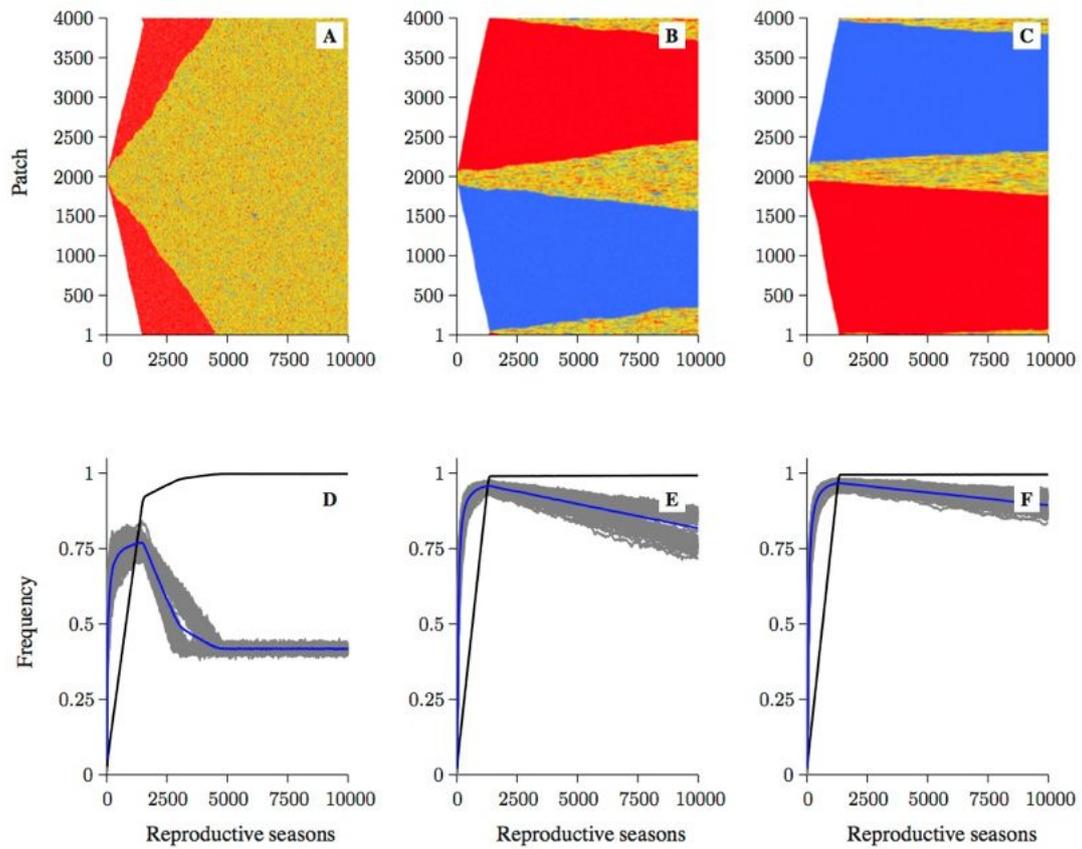

Figure S2. The importance of the short-range dispersal parameter $\alpha$. Here $\alpha = 6$, indicating increased short-range dispersal distance. For the explanation of the figure, refer to the caption of Fig. 2 in the main text. The relationship of sexual as compared to clonal reproduction is set to $c = 0.5$ (twice as costly to recruit asexual as compared to sexual offspring). Remaining parameters are the same as those used in Fig. 3.

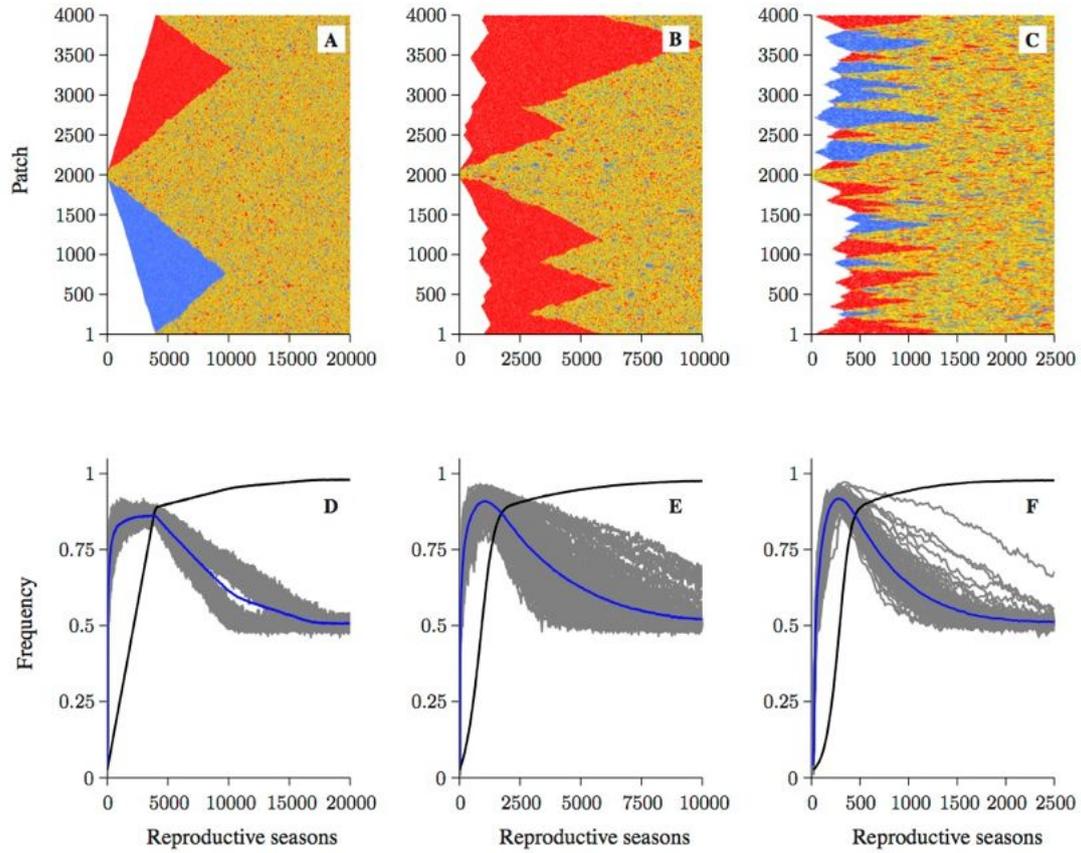

Figure S3. The importance of long-range dispersal. The values of the long-range dispersal rate are: $r = 0$ in A and D, $r = 10^{-4}$ in B and E, and $r = 10^{-3}$ in C and F. For the explanation of the results shown in different panels, refer to the caption of Fig. 2. The parameters in panel A correspond to those in Fig. 2B, but here we show the pattern obtained from a different stochastic realisation of the model. The results in panel D are the same as those shown in Fig. 2E. Remaining parameters used: $\beta = 1.25$, $N = 4000$, $\alpha = 3$, $d = 0.1$, $c = 0.5$, 100 stochastic realisations in D-F.

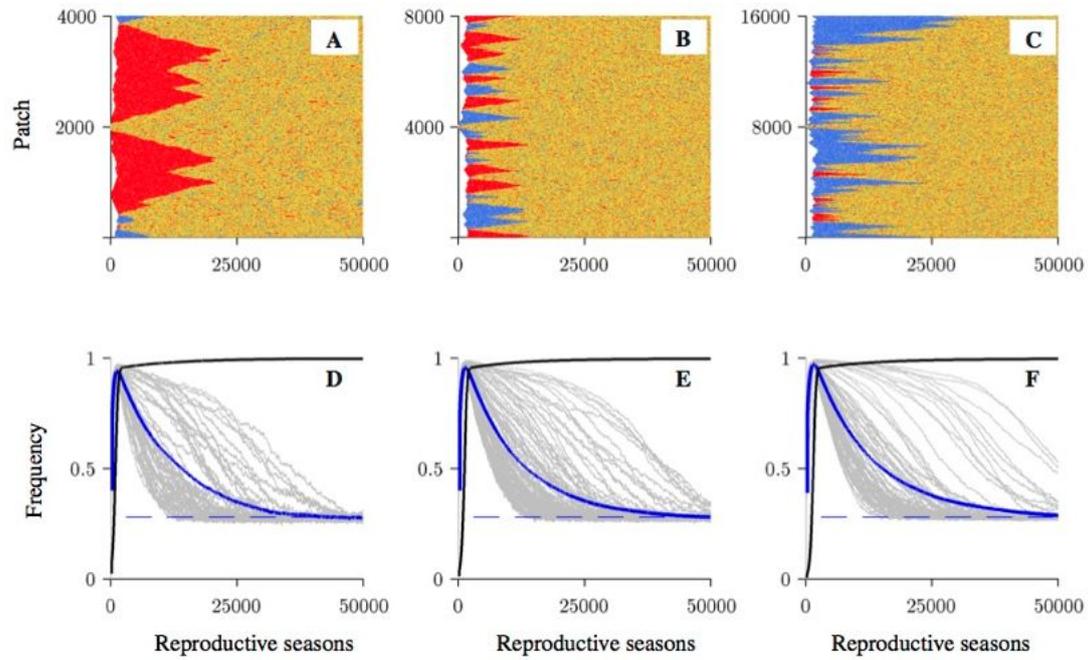

Figure S4. The importance of the habitat size. The habitat size is $N = 4000$ in A and D, $N = 8000$ in B and E, and $N = 16000$ in C and F. Dashed blue lines indicate the frequency of asexually recruited individuals in the quasi-steady state. For the explanation of remaining results shown in top and bottom panels refer to the caption of Fig. 2. Remaining parameters: $\beta = 1.25$, $\alpha = 3$, $d = 0.01$, $c = 0.2$, $r = 10^{-4}$, 100 stochastic realisations in D-F.

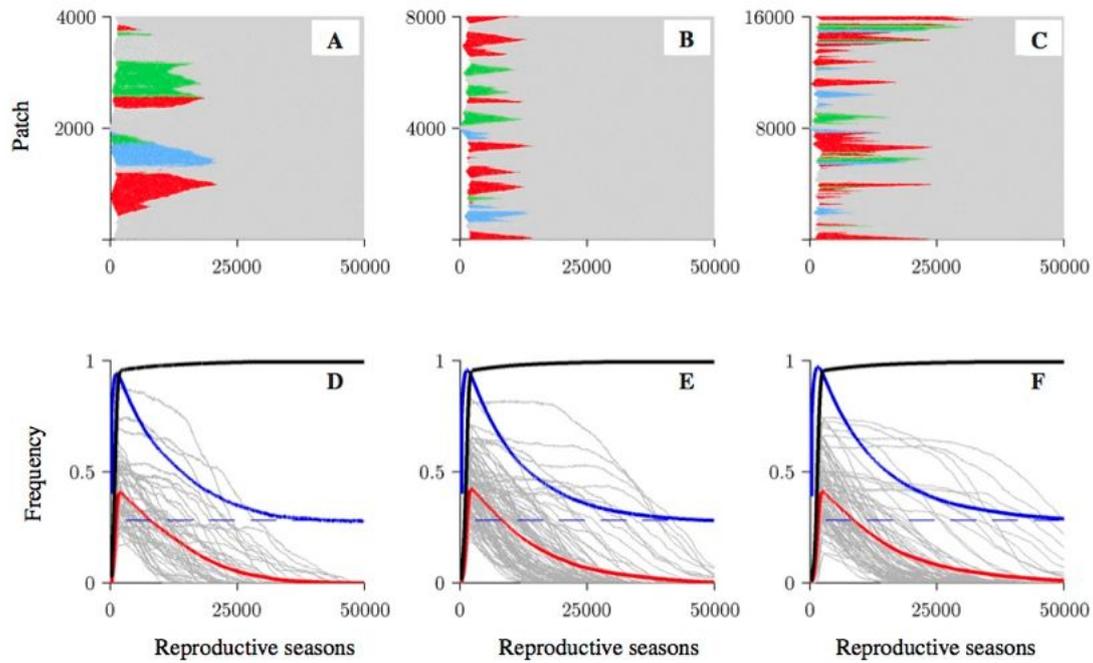

Figure S5. The importance of the habitat size for the distribution of dominant clones. The habitat size is $N = 4000$ in A and D, $N = 8000$ in B and E, and $N = 16000$ in C and F. Panels A-C show space-time patterns of three largest clonal colonies (coloured red, green, and blue) corresponding to the sexual patterns shown in Fig. S4A-C. Panels D-F show how the frequency of the largest dominant clone depends on time in 100 individual stochastic realisations (grey lines) for the parameters corresponding to those in panels A-C, respectively. Red lines are averages over the individual runs. Black, and solid and dashed blue lines are the same as those shown in Fig. S4D-F. For the remaining parameters refer to Fig. S4.

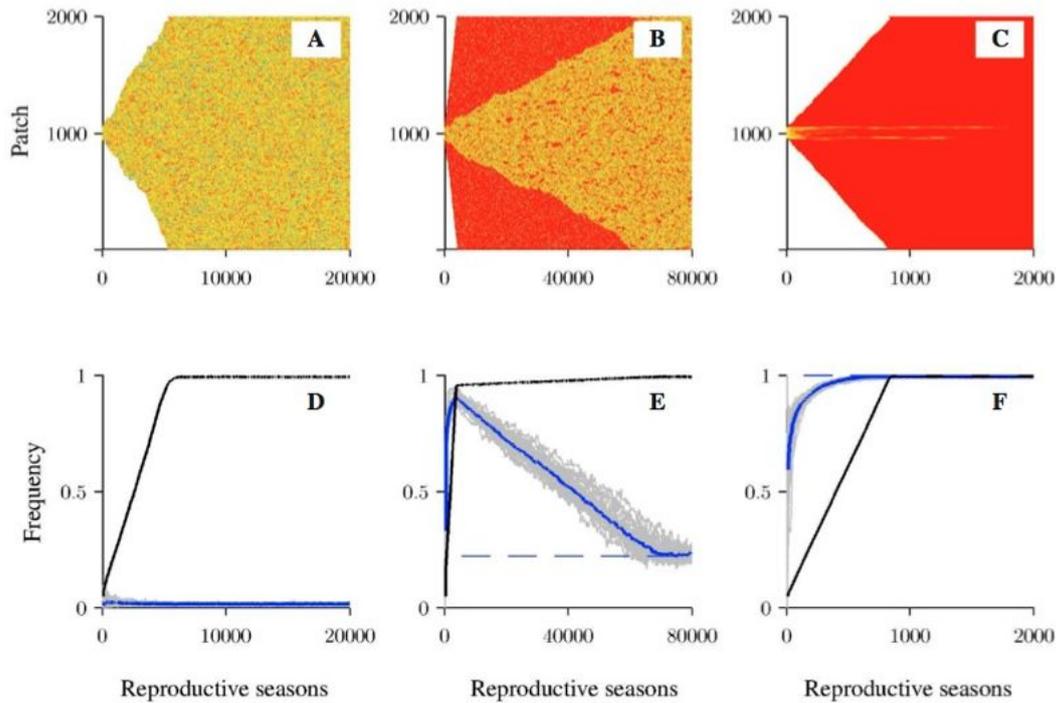

Figure S6. Sexual patterns in the model with males reproducing only sexually, and females reproducing both sexually and asexually. Panels A-C: space-time patterns of local sex ratios (pools of 8 neighbouring patches) obtained from single stochastic realisations of the model with the per-season per-female clonal birth rate $c = 0.02$ in A, $c = 0.2$ in B, and $c = 1$ in C. The per-season per-male clonal birth rate is zero. Empty patches are coloured white. Panels D-F: time dependence of the average frequency of occupied patches (black line), and of the average frequency of asexually recruited individuals (solid blue line) for the parameter values in A-C, respectively. Grey lines depict the results of 100 independent simulations. Dashed blue lines indicate the frequency of asexually recruited individuals in the quasi-steady state. Initial configuration: 100 neighbouring patches occupied, alternating sexes. The per-individual rate of production of sexual propagules is fixed to two. Remaining parameters: $N = 2000$, $\alpha = 3$, $d = 0.01$, $r = 0$.

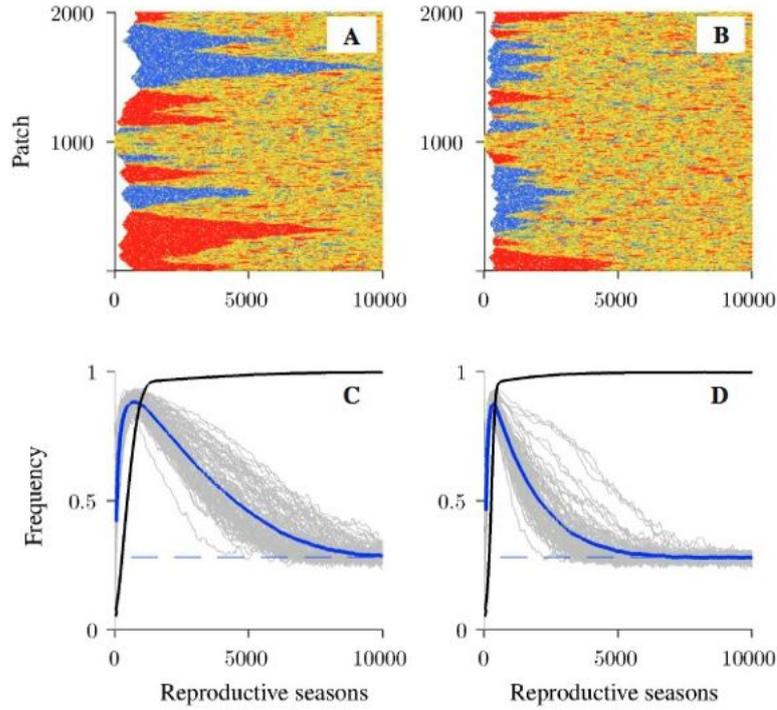

Figure S7. The effect of long-range dispersal of sexually produced seeds (A, C) as compared to long-range dispersal of adults or vegetative propagules (B, D) on the population sexual structure. Panels A, and B show space-time patterns of local sex ratios (pools of 8 neighbouring patches) obtained from single stochastic realisations of the model with long-range dispersal of seeds (A), and of adults (B). Blue depicts all males, and red all females. Empty patches are coloured white. Panels C, and D show how the average frequency of occupied patches (black line) and the average frequency of asexual recruits (solid blue line) depend on time (for the parameter values in A, and B, respectively). Grey lines depict the results of 100 independent simulations. Dashed blue lines indicate the frequency of asexually recruited individuals in the quasi-steady state. Initial configuration: 100 neighbouring patches populated, alternating sexes. The per-individual rate of production of sexual propagules is fixed to two. Remaining parameters: $N = 2000$, $\beta = 1.25$, $\alpha = 3$, $d = 0.01$, $r = 0.001$, $c = 0.2$.

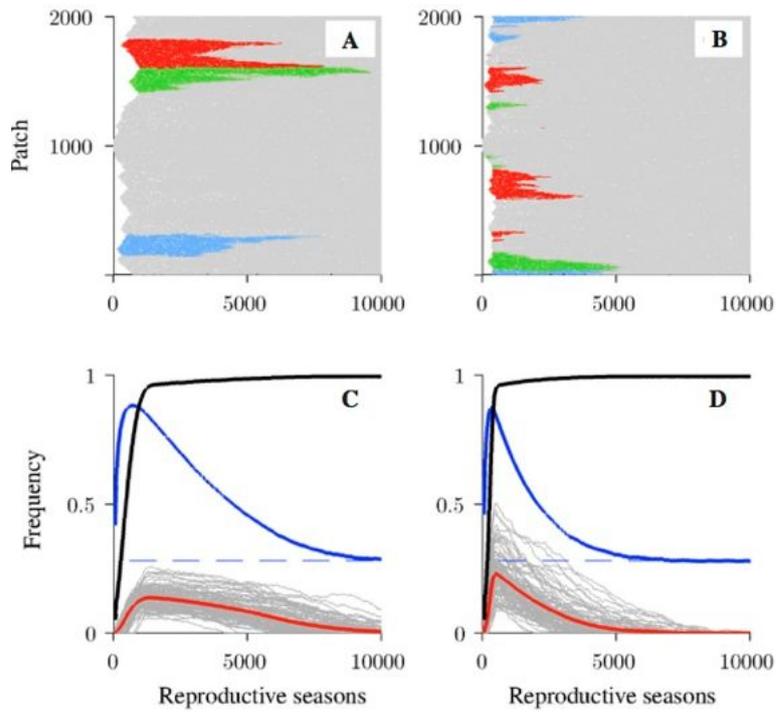

Figure S8. The effect of long-range dispersal of sexually produced seeds (A, C) as compared to long-range dispersal of adults or asexual propagules (B, D) on the dominant clones. Panels A, and B show space-time patterns of three largest clonal colonies (coloured red, green, and blue) corresponding to the sex patterns shown in Fig. S7A and Fig. S7B, respectively. Genotypes differing from the three largest clones are coloured grey, and empty patches are coloured white. Panels C-D show how the frequency of the largest dominant clone depends on time (grey lines, 100 independent realisations of the model). Red lines are averages over the individual runs. Dashed blue, and solid black and blue lines are the same as in Fig. S7. The initial configuration and the parameters used are the same as in Fig. S7.